\documentclass{article}
\usepackage{times}
\usepackage{graphics}
\def\rea{\noindent\hangafter=1\hangindent=1cm}

\def\deg{$^{\rm o}$}
\input epsf
\begin{document}
\renewcommand{\thefootnote}{\fnsymbol{footnote}}
\setcounter{footnote}{1}
\centerline{\large\bf Pulsating Stars in the ASAS-3 Database}
\centerline{\large\bf I. $\beta$~Cephei Stars}

\vspace{0.3cm}
\centerline{by}

\vspace{0.3cm}
\centerline{A.~P i g u l s k i}

\vspace{0.3cm}
\centerline{Instytut Astronomiczny Uniwersytetu Wroc{\l }awskiego,}
\centerline{Kopernika 11, 51-622 Wroc{\l }aw, Poland}
\centerline{E-mail: pigulski@astro.uni.wroc.pl}

\vspace{0.5cm}
\centerline{\it Received 2005 May 30}

\vspace{1cm}
\centerline{ABSTRACT}

\vspace{0.5cm}
{\small

We present results of an analysis of the ASAS-3 data for short-period variables from the recently published catalogue of 
over 38\,000 stars.  Using the data available in the literature we verify the results of the automatic classification
related to $\beta$~Cephei pulsators.  In particular, we find that 14 stars in the catalogue can be classified 
unambiguously as new $\beta$~Cephei stars.  By means of periodogram analysis we derive the frequencies and amplitudes of the 
excited modes.  The main modes in the new $\beta$~Cephei stars have large semi-amplitudes, 
between 35 and 80~mmag.   Up to four modes were found in some stars.  
Two (maybe three) new $\beta$ Cephei stars are members of southern young open clusters: ASAS\,164409$-$4719.1
belongs to NGC\,6200, ASAS\,164630$-$4701.2 is a member of Hogg\,22, and ASAS\,164939$-$4431.7 could be a member of NGC\,6216. 

We also analyze the photometry of four known $\beta$~Cephei stars in the ASAS-3 catalogue, namely IL\,Vel, NSV\,24078, 
V1449\,Aql and SY\,Equ.  Finally, we discuss the distribution of $\beta$~Cephei stars in the Galaxy.

\vspace{0.2cm} 

{\bf Keywords}: {\it stars: $\beta$ Cephei -- stars: pulsations -- stars: classification}
}
%%%

\vspace{0.5cm}
\centerline{\bf 1.~Introduction}

\vspace{0.5cm}
The idea of monitoring the whole sky for variability with small wide-field cameras, initiated and popularized by Pa\-czy\'n\-ski 
(1997, 2000, 2001),
resulted in several successful projects of which the All Sky Automated Survey (ASAS) conducted by Dr.~Grzegorz Poj\-ma\'n\-ski 
(Pojma\'nski 1997) is probably the most fruitful.  At the third stage of the project, ASAS-3, its ultimate goal was achieved: 
the whole southern sky and partly the northern sky were monitored for variability.  
The preliminary lists of variable stars resulting from the analysis of the photometric data obtained within the ASAS-3 project
have been recently
published in a series of four papers (Pojma\'nski 2002, 2003; Pojma\'nski and Maciejewski 2004, 2005) and are 
available in the Internet.
Hereafter, we will refer to the tables and data presented in these papers as the ASAS-3 catalogue.  The catalogue includes
over 38\,000 variable stars brighter than $\sim$14~mag in $V$. 

The $V$-band photometry of the ASAS-3 catalogue forms an excellent homogeneous source of data that can be used to study 
properties of different types of variable stars in the Galaxy.  In this series of papers, we will present the analysis
of the ASAS-3 photometry starting with the short-period early B-type pulsators, the $\beta$~Cephei stars. 

According to the newest review paper on $\beta$~Cephei stars (Stankov and Handler 2005), 93 certain objects
of this type are presently known in the Galaxy.  About half of them were found in young open clusters and OB associations; many, 
including the prototype, are naked-eye objects.  Typical periods range from 3 to 7 hours, the semi-amplitudes are usually smaller than
0.03~mag in $V$. Multiperiodic behavior is rather a rule than exception. The pulsations are identified with radial and nonradial
low-order $p$ and/or $g$ modes.

From the analysis of the ASAS-3 $V$-band photometry and the information available in the literature, we will show that
18 objects in the ASAS-3 database are {\it bona fide} $\beta$~Cephei stars, of which only four were previously known.  The remaining 
14 variable stars are new large-amplitude members of this interesting group of pulsators.  

\vspace{0.8cm}
\centerline{\bf 2.~Selection of Objects}

\vspace{0.5cm}
A preliminary selection and classification of variable stars were already done by the authors of the ASAS-3 catalogue.
Since several million stars were monitored during the ASAS-3 survey, fast algorithm for extracting variable objects
had to be implemented.  For this reason, the first selection of variables was made by the authors of the catalogue 
using the magnitude--dispersion diagram.  Only 
stars showing an excess of dispersion were selected as subjects of the subsequent periodogram analysis.  
The analysis allowed to find dominant periodicity reported in the catalogue.
Initially, the classification was made using Fourier coefficients, periods, amplitudes and visual inspection of the 
light curves (Poj\-ma\'nski 2002).  Starting with the second paper of the series (Poj\-ma\'nski 2003), 
the authors of the catalogue used additional information, i.e.,  
IRAS and 2MASS infrared photometry and galactic coordinates, to assign automatically the periodic variable stars to about a dozen
predefined classes.  This classification was only provisional, but could be very helpful in extracting objects of a given type.
The classification types were not exclusive, that is, many stars were assigned to more than one class.

The class BCEP, denoting $\beta$~Cephei stars, was included in the classification scheme in the third part of
the ASAS-3 catalogue (Pojma\'nski and Maciejewski 2004).  Eight BCEP stars were reported in that paper, another 14 were added in the
fourth paper (Pojma\'nski and Maciejewski 2005).  Adding four stars with positive declinations that appear only in the Internet
version of the ASAS-3 catalogue, ten stars with secondary or tertiary BCEP classification, and the known $\beta$~Cephei star IL~Vel 
classified as DSCT, denoting $\delta$~Scuti star, we get 37 stars that appear in the Internet list of $\beta$~Cephei 
stars in the ASAS-3 catalogue.
As we will show below, only 12 of them can be actually classified as {\it bona fide} $\beta$~Cephei stars.  
On the other hand, some $\beta$~Cephei 
stars could be included in the other classes, DSCT being the most probable one.  
The shortest periods observed in $\beta$~Cephei stars are of the order of 0.06~d, 
the median value for all known Galactic 
objects is about 0.17~d (Stankov and Handler 2005).  Periods of $\delta$~Scuti stars are, on the average, shorter.   
Their shortest periods are equal to about 0.02~d,
but there are $\delta$~Scuti stars with periods longer than 0.2~d (Rodr\'{\i}guez and Breger 2002).  
Both classes show multiperiodic behavior 
and small amplitudes.  Because of the wide range of overlapping periods, $\beta$~Cephei and $\delta$~Scuti stars cannot be 
distinguished solely from the observed periods.  
The only subgroup of $\delta$~Scuti stars that is relatively easy to indicate using periods and light curves
are the high-amplitude $\delta$~Scuti (HADS) stars.  They  
show large-amplitude variations, characteristic shape of the light curve and, typically, only one or two radial modes.

The easiest method to distinguish between $\beta$~Cephei and $\delta$~Scuti stars is using the information on their MK spectral types.
This is because $\beta$~Cephei stars cover a very narrow range of spectral types, between B0 and B2.5, while $\delta$ Scuti 
pulsators are A or early F-type stars.  Some photometric methods can also be used.  For example, the two classes separate well 
in the two-color 
($U-B$) vs.~($B-V$) diagram.  Unfortunately, the $UBV$ photometry is available only for some ASAS stars.  Pojma\'nski and Maciejewski 
(2004, 2005) used 
a homogeneous set of infrared measurements from the 2MASS survey 
as an additional filter in the classification.  In particular, they used ($J-H$) vs.~($H-K$) and $\log$(period) vs.~($J-H$) diagrams.
While some $\beta$~Cephei stars with small reddenings can be separated from $\delta$~Scuti objects in this way, 
this method fails for more reddened stars.

In this situation, we decided to proceed in the following way.  First, we analyzed the ASAS-3 photometry of 
all 37 stars that were classified as BCEP in the ASAS-3 catalogue.  We also checked their $UBV$ photometry (if available), 
H$\beta$ photometry and spectral types.  In this way, we found that 12 stars from this sample
can be reliably classified as $\beta$~Cephei pulsators.  Four of them were already known as stars of this type.
Furthermore, we checked the available MK spectral types and $UBV\beta$ photometry for all stars in the ASAS-3 catalogue that
had: (i) full amplitude, $\Delta V$, smaller than 0.3~mag (note that BW Vul, $\beta$~Cephei star with the exceptionally large amplitude
has $\Delta V \approx $ 0.2~mag), and (ii) dominating period shorter than 0.5~d 
(for all classes of pulsating stars as well as the variables classified as ACV and MISC) 
or 1~d (for three classes of eclipsing binaries: EC, ESD, ED).  There were about 1700 stars in the ASAS-3 catalogue satisfying 
both criteria.  As a result, we found six additional $\beta$~Cephei stars.  
All eighteen stars of this type found in the ASAS-3 database are listed in Table 1.

%
%     =========  Table 1 ==============
%
\begin{center}
{\small
T a b l e\quad 1\\
$\beta$ Cephei stars in the ASAS-3 database

\vspace{2mm}
\begin{tabular}{rcclrll}
\hline\noalign{\smallskip}
 HD & CPD/BD & ASAS name& ASAS classification & \multicolumn{1}{c}{$V$}   
 & MK sp.~type & Notes \\
\noalign{\smallskip}\hline\noalign{\smallskip}
 80383 & $-$52{\deg}2185 & 091731--5250.3 & DSCT &  9.14 & B2\,III (1) & IL Vel \\
164340 & $-$40{\deg}8357 & 180233--4005.2 & BCEP/EC/ESD & 9.28 & B0\,III (2) & NSV\,24078\\
180642 & $+$00{\deg}4159 & 191715+0103.6 & BCEP & 8.27 & B1.5\,II-III (3) & V1449\,Aql\\
203664 & $+$09{\deg}4973 & 212329+0955.9 & BCEP/EC/ESD & 8.57 & B0.5\,IIIn (3) & SY Equ\\
\noalign{\smallskip}\hline\noalign{\smallskip}
 ---   & $-$62{\deg}2707 & 122213--6320.8 & BCEP & 10.06 & B2\,III (4) & \\
133823 & $-$65{\deg}2993 & 150955--6530.4 & BCEP &  9.62 & B2\,IV (4) & \\
 ---   & $-$50{\deg}9210 & 161858--5103.5 & BCEP/EC & 10.33 & B2\,II (5) & \\
328862 & $-$47{\deg}7861 & 164409--4719.1 & BCEP/DSCT & 10.13 & B0.5\,III (6) & in NGC\,6200\\
 ---   & $-$46{\deg}8213 & 164630--4701.2 & RRC/DSCT/EC/ESD & 10.86 & --- & in Hogg~22\\
328906 & --- & 164939--4431.7 & DSCT/EC/ESD & 11.22 & [B2] (7)& in NGC\,6216? \\
152077 & $-$43{\deg}7731 & 165314--4345.0 & BCEP/DSCT &  9.08 & B1 II (4) & \\
152477 & $-$47{\deg}7958 & 165554--4808.8 & ESD/RRC/EC &  9.03 & B1 II (4) & \\
155336 & $-$32{\deg}4389 & 171218--3306.1 & BCEP/DSCT &  9.46 & B1/2 Ib (8) & \\
165582 & $-$34{\deg}7600 & 180808--3434.5 & BCEP & 9.39 & B1\,II (4) & \\
167743 & $-$15{\deg}4909 & 181716--1527.1 & BCEP=DSCT & 9.59 & B2\,Ib (9) & \\
 ---   &  ---  & 182610--1704.3 & DSCT & 10.21 & --- & ALS\,5036\\
 ---   &  ---  & 182617--1515.7 & DSCT & 10.73 & --- & ALS\,5040\\
 ---   & $-$14{\deg}5057 & 182726--1442.1 & EC/ESD & 9.99 & --- & \\

\noalign{\smallskip}\hline
\end{tabular}

\vspace{2mm}

}
\end{center}
{\small References to MK spectral types in Table 1:
(1) Houk (1978), (2) Hill {\it et al.}~(1974), (3) Walborn (1971), (4) Garrison {\it et al.}~(1977), (5) FitzGerald (1987), 
(6) Whiteoak (1963), (7) Spectral type on the Harvard system, Nesterov {\it et al.}~(1995), (8) Houk (1982), 
(9) Houk and Smith-Moore (1988).
}

\vspace{0.8cm}
\centerline{\bf 3.~The Analysis}

\vspace{0.5cm}
The ASAS-3 data for a given star consist of the aperture photometry made through five different apertures.  Since both 
magnitudes and their errors are reported in each aperture, we chose for analysis the data made in the aperture 
which had the smallest mean error.  The photometry for a star in the catalogue comprises typically 
several independent subsets of measurements, as the star was usually observed in several different fields.
These subsets may differ in the mean magnitude, as pointed out by the authors of the catalogue.
In addition, a quality grade (from A to D) was assigned to each measurement.  
The procedure of analyzing the data was the following.  First, the individual subsets were extracted.  
We used only the measurements that were flagged A or B.  
Because of the magnitude offsets between subsets, the mean magnitude was subtracted from each
subset and then the subsets were combined.  These combined data were subject of preliminary periodogram analysis.  

After subtracting all significant signals from the combined data, the residuals were checked and the following operations were 
performed: (i) The outliers were rejected from the original data using 3-$\sigma$ clipping in residuals.
(ii) The mean offset for each subset was calculated from the residuals and, along 
with the mean magnitude derived earlier, subtracted from the original data.  (iii) The long-term trend was removed from the data.  This
was done by calculating average residuals in 200-day intervals
that were used to derive a smoothed curve with a cubic spline fit.  This smoothed curve was removed from the original data.  

The combined, cleaned and detrended data were used in the final analysis.  The analysis included: (i) calculating
Fourier periodogram in the range from 0 to 30~d$^{-1}$, (ii) fitting a sinuoid with the frequency of the highest peak, as well as all
frequencies found earlier, (iii) improving frequencies, amplitudes and phases by means of the non-linear least-squares method.  The 
residuals from the fit were used in the next iteration as the input file.  The extraction of frequencies was performed until the 
signal-to-noise ratio (S/N) in the periodogram became smaller than 4.  However, some frequencies with S/N barely exceeding 4 were not 
included in the final solution.  By the final solution we mean the results of fitting the data with a series 
of sinusoidal terms in the form of
\begin{equation}
\sum\limits_{i=1}^{n} A_i\sin[2\pi F_i(t-T_0) + \phi_i], 
\end{equation}
where
\begin{equation}
F_i = \sum\limits_{j=1}^{k} m_j f_j,
\end{equation}
is a linear combination of independent frequencies, $f_j$, extracted from the periodograms. This form was used to 
allow the combination frequencies and harmonics found in the periodograms to be fitted. 
In the above equations, $A_i$ denote 
semi-amplitude, $t$ is the time reckoned from the initial epoch $T_0$ = HJD\,2450000.0, $\phi_i$ is the phase, and $m_j$ is an integer.
The parameters of the fits are given in Table 2 for all 18 $\beta$~Cephei stars listed in Table 1.  However, instead of listing
phases $\phi_i$, we provide the times of maximum light, $T^i_{\rm max} - T_0$, closest to the mean epoch of all observations.
It Table 2, $\sigma_{\rm res}$ denotes standard deviation of the residuals, DT is the detection threshold corresponding to S/N = 4.
The other columns are self-explanatory.  The r.m.s.~errors of the last digits are given in parentheses.

The Fourier 
periodograms showing the consecutive steps of prewhitening are shown in Figs.~1--4 
and commented in the next section.  As can be judged from these figures, the daily aliases in the periodograms of the 
ASAS-3 data are quite strong.  Consequently, there is some ambiguity in the derived frequencies, especially for the low-amplitude terms.

\vspace{0.8cm}
\centerline{\bf 4.~Notes on Individual Stars}

\vspace{0.5cm}
{\it 4.1.~Known $\beta$~Cephei-Type Stars}

\vspace{0.3cm}
IL Vel = HD\,80383 = CPD\,$-$52$^{\rm o}$2185, B2\,III.  The pulsations of IL~Vel were discovered by Haug (1977, 1979).  
The most extensive photometry was obtained
by Heynderickx and Haug (1994) and Handler {\it et al.}~(2003).  The latter authors showed convincingly that the pulsational
spectrum of IL~Vel is dominated by two large-amplitude modes closely spaced in frequency.  Handler {\it et al.}~(2003) found a third 
mode, with a much lower amplitude.  We detected the two large-amplitude modes in the ASAS-3 data (Fig.~1).  Our frequencies  
(see Table 2) agree with those of Handler {\it et al.}~(2003), but the amplitudes are $\sim$10\% smaller.
The $V$ amplitude of the third mode found by Handler {\it et al.}~(2003) amounted to 6.5 $\pm$ 0.7~mmag.  We do not detect 
this mode despite the fact that the detection threshold for this star in the ASAS-3 data is equal to 4.8~mmag.

NSV\,24078 = HD\,164340 = CPD\,$-$40$^{\rm o}$8357 = HIP\,88352, B0\,III. The star was suspected to be variable from the 
{\it Hipparcos} data.  
The data from this satellite were analyzed by Molenda-\.Zakowicz (private communication), who found two
periodicities.  Consequently, the star was included into the catalogue of Stankov and Handler (2005). 
We detected the same two frequencies in the ASAS-3 data (Fig.~1).  There is an indication of a third frequency 
($f_3 \approx$ 3.9278~d$^{-1}$ or one of the daily aliases), but 
the S/N for this frequency
equals to about 4.5, that is, it only slightly exceeds the adopted detection level 
(S/N = 4).
We therefore did not include this frequency in the final solution.

V1449~Aql = HD\,180642 = ALS\,10235 = HIP\,94793, B1.5\,II-III. This star was found to be 
variable in the {\it Hipparcos} data (Waelkens {\it et al}.~1998, Aerts 2000).   A single frequency of 5.4871~d$^{-1}$ and $V$ amplitude of 39~mmag
was found by Aerts (2000).  This frequency with almost the same amplitude is confirmed in the ASAS-3 data (Table 2, Fig.~1).  
No other periodicities with an amplitude exceeding 6~mmag were found.

SY Equ = HD 203664 = BD\,+9$^{\rm o}$4973, B0.5\,IIIn.  This is another star found by {\it Hipparcos} and reobserved by Aerts (2000).  
She found a single mode with frequency of 6.0289~d$^{-1}$ in the {\it Hipparcos} data.  We confirm this frequency
from the ASAS-3 observations (Fig.~1, Table 2).

\begin{figure}[!ht]
\epsfbox{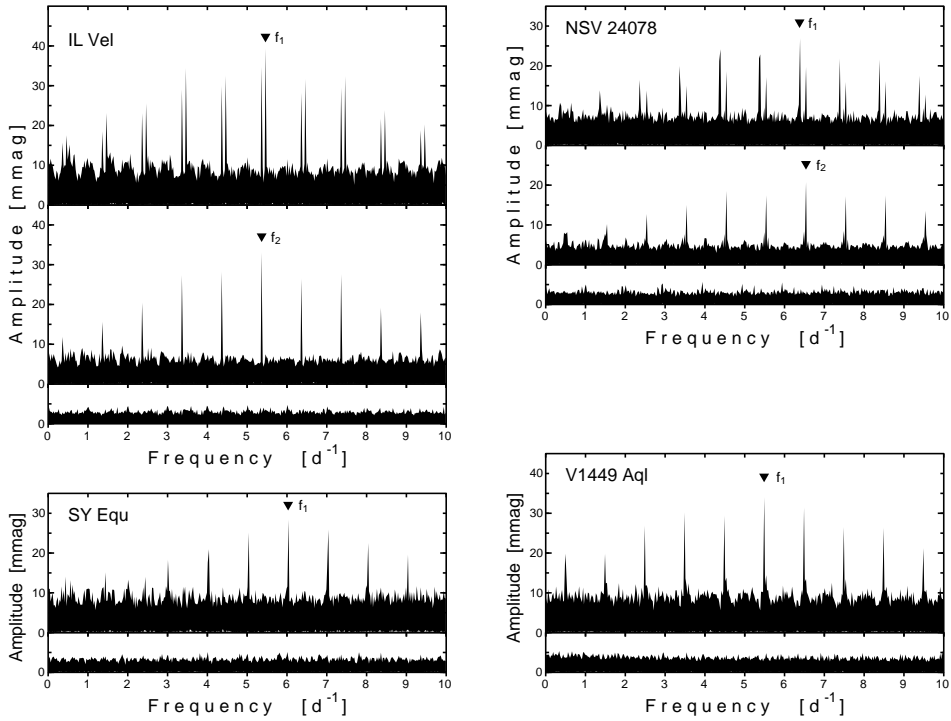}
\caption{\small Fourier frequency spectra showing the consecutive steps of prewhitening for the four known $\beta$~Cephei stars 
in the ASAS-3 catalogue: IL~Vel, NSV\,24078, V\,1449~Aql, and SY~Equ.} 
\label{four1}
\end{figure}

\vspace{0.5cm}
{\it 4.2.~New $\beta$~Cephei-Type Stars}

\vspace{0.3cm}
CPD\,$-$62$^{\rm o}$2707 = ALS\,2653.  The MK spectral type of this star (B2\,III) was given by Garrison {\it et al.}~(1977).  The $UBV$ 
photometry of Schild {\it et al.}~(1983) and the $UBV\beta$ photometry of Klare and Neckel (1977, hereafter KN77) indicate that 
it is indeed
an early B-type star.  The pulsation spectrum consists of a large-amplitude single mode with frequency $f_1$ = 7.0589~d$^{-1}$
whose harmonic, 2$f_1$, is also detected in the ASAS-3 data (Fig.~2).

HD\,133823 = CPD\,$-$65$^{\rm o}$2993.  Two MK spectral types are reported for this star in the literature.  
Houk and Cowley (1975) classify HD\,133823 as a B3\,II star, while Garrison {\it et al.}~(1977) give B2\,IV.  
We find a single frequency of $f_1$ = 5.6804~d$^{-1}$ in the ASAS-3 data.  A significant signal at 0.4989~d$^{-1}$ appears
in the periodogram after removing $f_1$ (Fig.~2), but it is likely to be spurious due to its proximity to 0.5~d$^{-1}$.

CPD\,$-$50$^{\rm o}$9210 = ALS\,3547.  The MK spectral type given by FitzGerald (1987) is B2\,II.  The available
$UBV$ photometry (FitzGerald 1987, Reed and Vance 1996) are also typical for an early B-type star.  We detect two close frequencies 
in the ASAS-3 data (Fig.~2, Table 2).

\vfill\eject
%===========================
%   Table 2: beta Cep stars
%===========================
\begin{center}
{\small

 T a b l e\quad 2\\
Parameters of the sine-curve fits to the $V$ 
magnitudes of the $\beta$~Cephei stars in the ASAS-3 database \\

\vspace{2mm}
\begin{tabular}{cccrrccr}
\hline\noalign{\smallskip}
Star & $f$ & $N_{\rm obs}$ & \multicolumn{1}{c}{$f_i$} & $A_i$ & 
$T_{\rm max}^i - T_0$   &  $\sigma_{\rm res}$ / DT \\  
   &     &  &\multicolumn{1}{c}{[d$^{-1}$]} & [mmag] 
& [d] & [mmag] \\
\noalign{\smallskip}\hline\noalign{\smallskip}
IL\,Vel & $f_1$ & 367 & 5.459781(08) & 37.9(09) & 2741.3991(07) & 12.8 / 4.8 \\
        & $f_2$ &     & 5.363255(09) & 34.7(10) & 2741.4018(08) &            \\
\noalign{\smallskip}\hline\noalign{\smallskip}
NSV\,24078& $f_1$ & 270 & 6.377727(14) & 27.5(10) & 2673.2062(09) & 11.7 / 5.0 \\
          & $f_2$ &     & 6.538740(18) & 20.9(10) & 2673.2280(12) &            \\
\noalign{\smallskip}\hline\noalign{\smallskip}
V\,1449 Aql& $f_1$ & 182 & 5.486928(14) & 36.8(12) & 2854.3232(09) & 11.3 / 5.9 \\
\noalign{\smallskip}\hline\noalign{\smallskip}
SY Equ& $f_1$ & 106 & 6.028753(24) & 29.7(12) & 2893.1065(10) & 8.5 / 5.8 \\
\noalign{\smallskip}\hline\hline\noalign{\smallskip}
CPD\,$-$62$^{\rm o}$2707 & $f_1$ & 380 & 7.058920(05) & 55.1(10) & 2695.1646(04) & 13.9 / 5.1 \\
                         & 2$f_1$ & &14.117840~~~~~~~ & ~~9.0(10) & 2695.1675(13)  &\\
\noalign{\smallskip}\hline\noalign{\smallskip}
HD\,133823 & $f_1$ & 287 & 5.680437(10) & 52.9(15) & 2692.0050(08) & 18.2 / 7.5 \\
\noalign{\smallskip}\hline\noalign{\smallskip}
CPD\,$-$50$^{\rm o}$9210 & $f_1$ & 280 & 4.866859(13) & 39.5(15) & 2687.1942(13) & 17.3 / 7.4 \\
                         & $f_2$ &     & 4.879365(27) & 19.4(14) & 2687.1390(25) & \\
\noalign{\smallskip}\hline\noalign{\smallskip}
HD\,328862 & $f_1$ & 262 & 4.948815(07) & 82.7(17) & 2701.9660(07) & 19.2 / 8.4 \\
           & $f_2$ &     & 4.924633(34) & 15.1(16) & 2701.9864(38) & \\
           & $f_3$ &     & 5.398964(49) & 11.6(18) & 2701.9490(42) & \\
     & $f_1 + f_2$ &     & 9.873448~~~~~~~   & 10.5(18) & 2701.9762(26) & \\
\noalign{\smallskip}\hline\noalign{\smallskip}
CPD\,$-$46$^{\rm o}$8213 & $f_1$ & 232 & 4.460172(13) & 64.4(23) & 2723.1942(13) & 25.3 / 11.8 \\
\noalign{\smallskip}\hline\noalign{\smallskip}
HD\,328906 & $f_1$ & 256 & 5.630744(23) & 41.8(28) & 2679.1349(19) & 31.8 / 14.1 \\
\noalign{\smallskip}\hline\noalign{\smallskip}
HD\,152077 & $f_1$ & 405 & 4.911496(10) & 51.0(13) & 2584.6736(08) & 17.7 /  6.4 \\
           & $f_2$ &     & 4.851653(17) & 25.0(12) & 2584.5983(17) & \\
           & $f_3$ &     & 4.886405(24) & 18.5(13) & 2584.6192(22) & \\
      & $f_1 + f_2$&     & 9.763149~~~~~~~~& 9.1(13)& 2584.7367(22) & \\
           & $f_4$ &     & 3.985732(66) &  7.7(13) & 2584.6873(65) & \\
\noalign{\smallskip}\hline\noalign{\smallskip}
HD\,152477 & $f_1$ & 313 & 3.773723(10) & 35.4(10) & 2702.3268(12) & 12.5 / 5.0 \\
\noalign{\smallskip}\hline\noalign{\smallskip}
HD\,155336 & $f_1$ & 620 & 5.531762(08) & 46.5(08) & 3007.3924(05) & 13.7 / 5.6 \\
           & $f_2$ &     & 5.400986(32) & 10.6(08) & 3007.4848(23) & \\
           & $f_3$ &     & 3.980814(38) &  8.5(08) & 3007.3636(37) & \\
\noalign{\smallskip}\hline\noalign{\smallskip}
HD\,165582 & $f_1$ & 222 & 4.747411(23) & 38.9(18) & 2874.1057(15) & 18.0 / 9.0 \\
           & $f_2$ &     & 7.390277(40) & 22.5(18) & 2874.0508(17) & \\
           & $f_3$ &     & 4.725015(50) & 16.3(18) & 2874.1073(37) & \\
           & $f_4$ &     & 3.381100(67) & 15.0(17) & 2874.2119(59) & \\
     & $f_1 + f_3$ &     & 9.472426~~~~~~~& 10.1(17)& 2874.1081(31)& \\
\noalign{\smallskip}\hline\noalign{\smallskip}
HD\,167743 & $f_1$ & 300 & 4.823737(10) & 41.6(12) & 2662.1794(09) & 14.2 / 5.8 \\
           & $f_2$ &     & 5.096927(17) & 26.4(12) & 2662.1918(14) & \\
           & $f_3$ &     & 4.975822(30) & 13.8(13) & 2662.1817(26) & \\
\noalign{\smallskip}\hline\noalign{\smallskip}
ALS\,5036  & $f_1$ & 520 & 4.917343(08) & 56.5(10) & 2722.9005(06) & 15.2 / 4.9 \\
           & $f_2$ &     & 4.919087(33) & 13.3(10) & 2722.8908(24) & \\
\noalign{\smallskip}\hline\noalign{\smallskip}
ALS\,5040  & $f_1$ & 365 & 4.973861(14) & 50.0(17) & 2695.9661(11) & 22.2 / 8.5 \\
           & $f_2$ &     & 5.071912(35) & 20.2(17) & 2695.9670(26) & \\
           & $f_3$ &     & 4.523109(69) & 10.6(17) & 2696.0228(56) & \\
\noalign{\smallskip}\hline\noalign{\smallskip}
BD\,$-$14$^{\rm o}$5057 & $f_1$ & 323 & 4.163681(09) & 43.9(10) & 2674.0466(09) & 12.7 / 5.0 \\
\noalign{\smallskip}\hline

\end{tabular}

}
\end{center}

\vfill\eject
HD\,328862 = CPD\,$-$47$^{\rm o}$7861 = ALS\,3721 = NGC\,6200\,\#4.  The MK spectral type reported by Whiteoak (1963) is B0.5\,III, while
FitzGerald {\it et al.}~(1977) classified the star as B1\,III:.  We detect three frequencies in the ASAS-3 data (Table 2, Fig.~2)
and the combination term, $f_1 + f_2$.
(Table 2).  The amplitude of the main mode is the largest among all stars reported in this paper and is comparable to that of BW\,Vul.
The star is a member of the loose open cluster NGC\,6200 in Ara (FitzGerald {\it et al.}~1977).
\begin{figure}[!ht]
\epsfbox{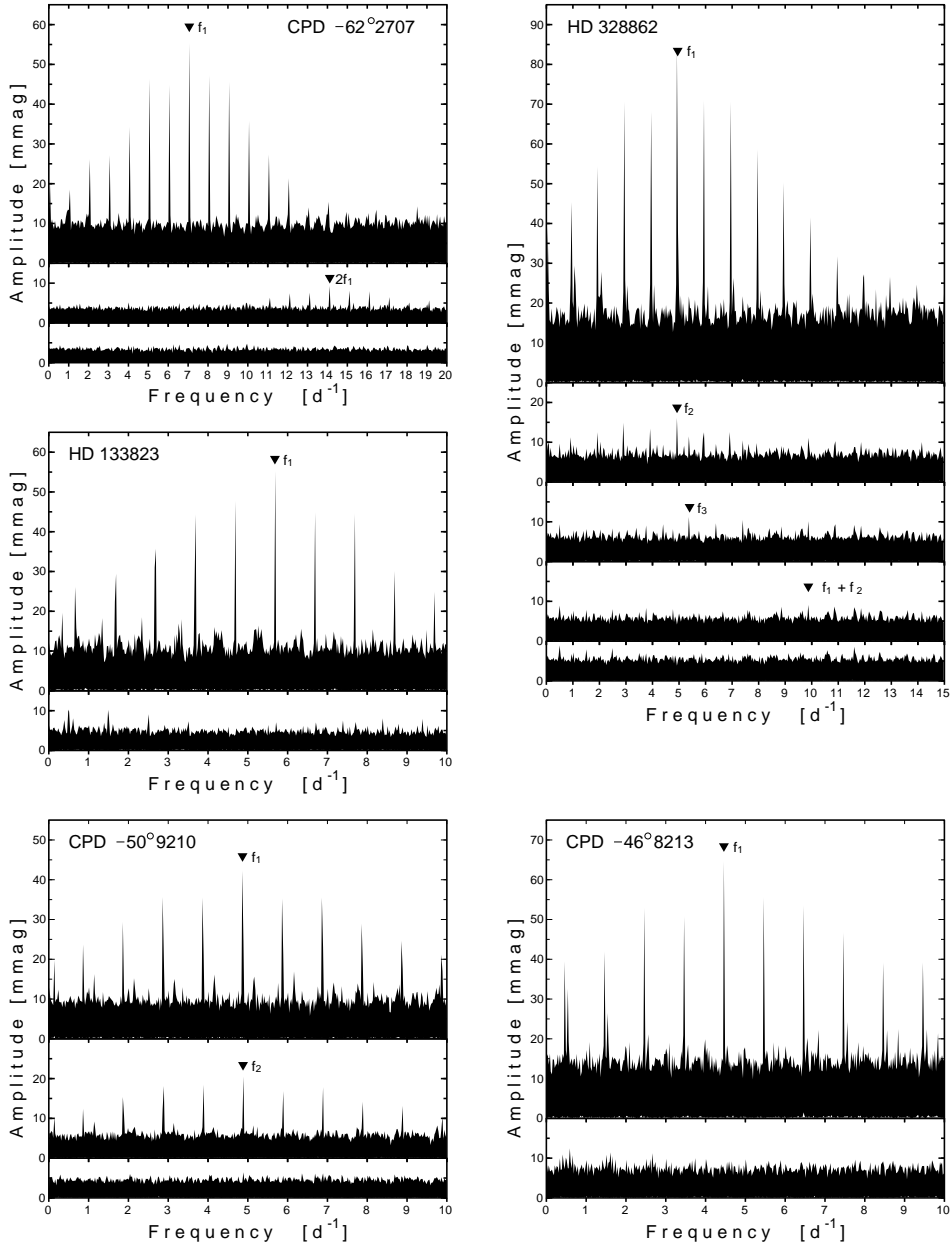}
\caption{\small The same as in Fig.~1, but for five new $\beta$~Cephei stars: CPD\,$-$62$^{\rm o}$2707, HD\,133823, 
CPD\,$-$50$^{\rm o}$9210, HD\,328862, and CPD\,$-$46$^{\rm o}$8213.}
\label{four2}
\end{figure}

CPD\,$-$46$^{\rm o}$8213 = Hogg\,22\,\#67.  No MK spectral type is available for this star.
However, from the $UBV$ photometry of Forbes and Short (1996) we may conclude that it is a reddened early B-type 
star and a likely member of a very young open cluster Hogg\,22 (Hogg 1965).  
In the sky, the cluster is located 6$^\prime$ off NGC\,6204, another
open cluster, but is twice as distant, much younger and more reddened than NGC\,6204 (Whiteoak 1963, 
Moffat and Vogt 1973, Forbes and Short 1996).  
The age of Hogg\,22 was estimated by Forbes and Short (1996) to be 5 $\pm$ 2~Myr.
In the ASAS-3 data of CPD\,$-$46$^{\rm o}$8213 we detect a single mode
with a frequency of about 4.4602~d$^{-1}$ (Fig.~2).
\begin{figure}[!ht]
\epsfbox{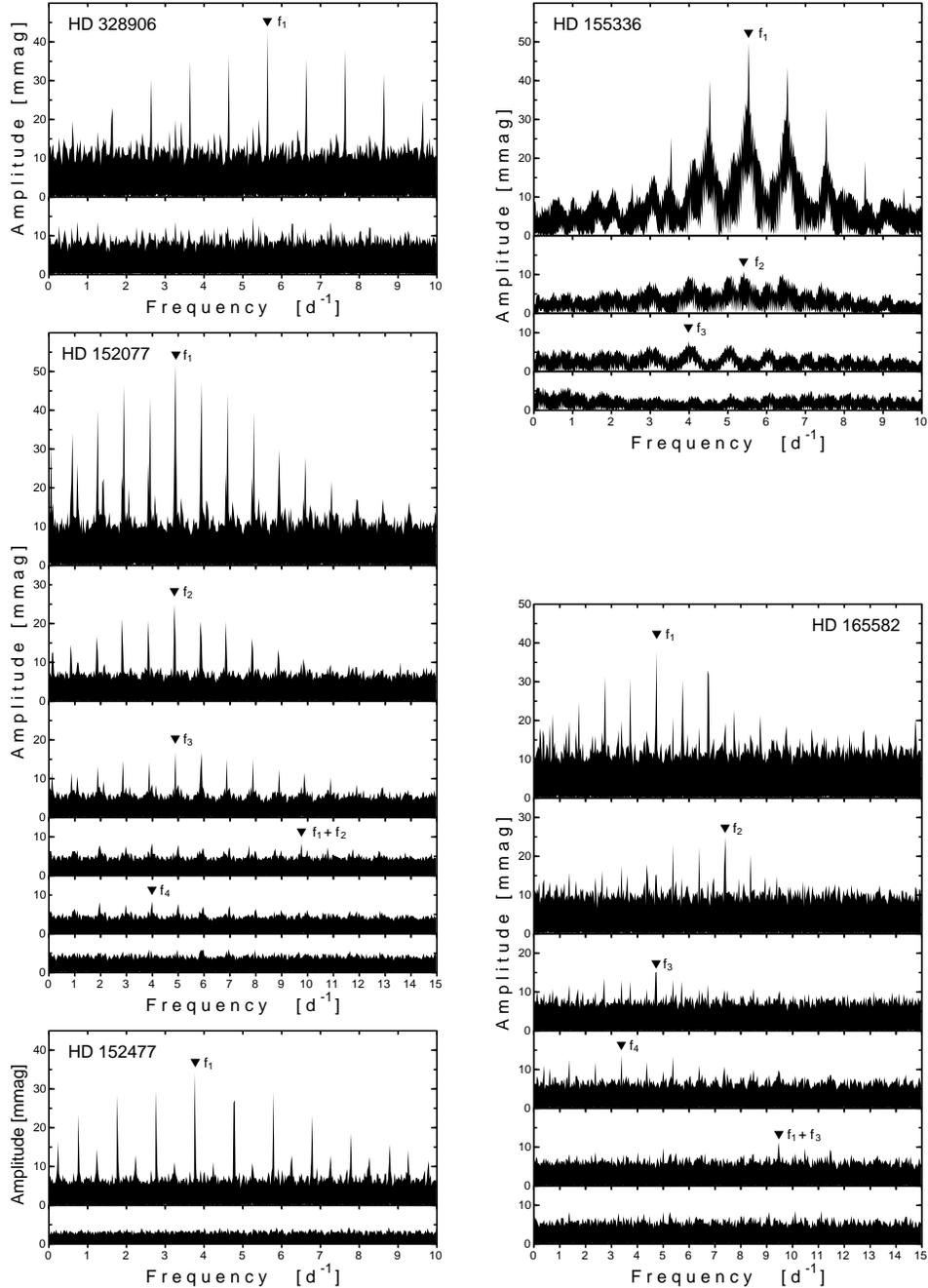}
\caption{\small The same as in Fig.~1, but for HD\,328906, HD\,152077, HD\,152477, HD\,155336, and HD\,165582.}
\label{four3}
\end{figure}

HD\,328906 = CD\,$-$44$^{\rm o}$11167.  Neither MK spectral type not the $UBV$ photometry is available for HD\,328906.  However, 
it has the Harvard spectral type of B2 (Nesterov {\it et al.}~1995). This allows us to classify the star
as a new $\beta$~Cephei variable.
We detect a mode with frequency $f_1 \approx$ 5.6307~d$^{-1}$ in the ASAS-3 data. There is an indication of the presence of
a second mode, at frequency 5.2587~d$^{-1}$ (Fig.~3),
as it stands slightly above the adopted detection threshold.  The star is located 
about 12$^\prime$ north of the young open cluster NGC\,6216.  The age of the
cluster is 35 $\pm$ 15~Myr (Piatti {\it et al.}~2000), still young enough to contain a $\beta$~Cephei star. However,
the membership of HD\,328906 has to be verified.

HD\,152077 = CPD\,$-$43$^{\rm o}$7731 = ALS\,3793.  According to Houk (1978), the MK spectral type is B2\,Iab/Ib, but
Garrison {\it et al}.~(1977) give B1\,II.  
However, the Str\"omgren $\beta$ index measured by KN77 and Knude (1992), amounting to 2.593 and 2.610, respectively,
is rather too large for a supergiant.   The $UBV$ photometry is available from KN77, 
Dachs {\it et al}.~(1982) and Schild {\it et al}.~(1983).
The pulsational spectrum of HD\,152077 is quite rich; we detect four independent 
modes including a triplet 
near 4.9~d$^{-1}$ (Fig.~3).  The triplet is, however, not equidistant in frequency as one would expect for a rotationally split mode.
The difference $f_1 - f_3$ = 0.02509 $\pm$ 0.00003~d$^{-1}$, while $f_3 - f_2$ =
0.03475 $\pm$ 0.00003~d$^{-1}$.  The fourth detected frequency, $f_4$ = 3.9857~d$^{-1}$, is far from the triplet, but it is quite 
likely that it is the 1~d$^{-1}$ alias at 4.9884~d$^{-1}$ that is the correct frequency.  
We also detect a combination frequency, $f_1 + f_2$.

HD\,152477 = CPD\,$-$47$^{\rm o}$7958.  The star was classified as B1\,Ib by Houk (1978) and B1\,II by Garrison {\it et al.}~(1977).
The $UBV$ photometry was obtained by KN77 and Schild {\it et al.}~(1983).  KN77 measured also its
$\beta$ index to be 2.608.  We find a single mode with frequency 3.7737~d$^{-1}$ 
(Fig.~3).

HD\,155336 = CPD\,$-$32$^{\rm o}$4389 = ALS\,3961.  Roslund (1966) and Houk (1982) give B3\,III and B1/B2 Ib, respectively. 
The $UBV$ photometry was first reported by Roslund (1964) and $UBV\beta$ photometry, 
by KN77.  We find three modes in the ASAS-3 data of HD\,155336.  As 
for HD\,152077, we suspect that the 1~d$^{-1}$ alias of $f_3$, at 4.9808~d$^{-1}$, may be the correct value.  The ASAS-3 data for
this star are distributed less evenly than for the other ones, so that the alias structure is also different.  This is because 
the field with this star was observed more frequently during six nights around JD\,2453200.
\begin{figure}[!ht]
\epsfbox{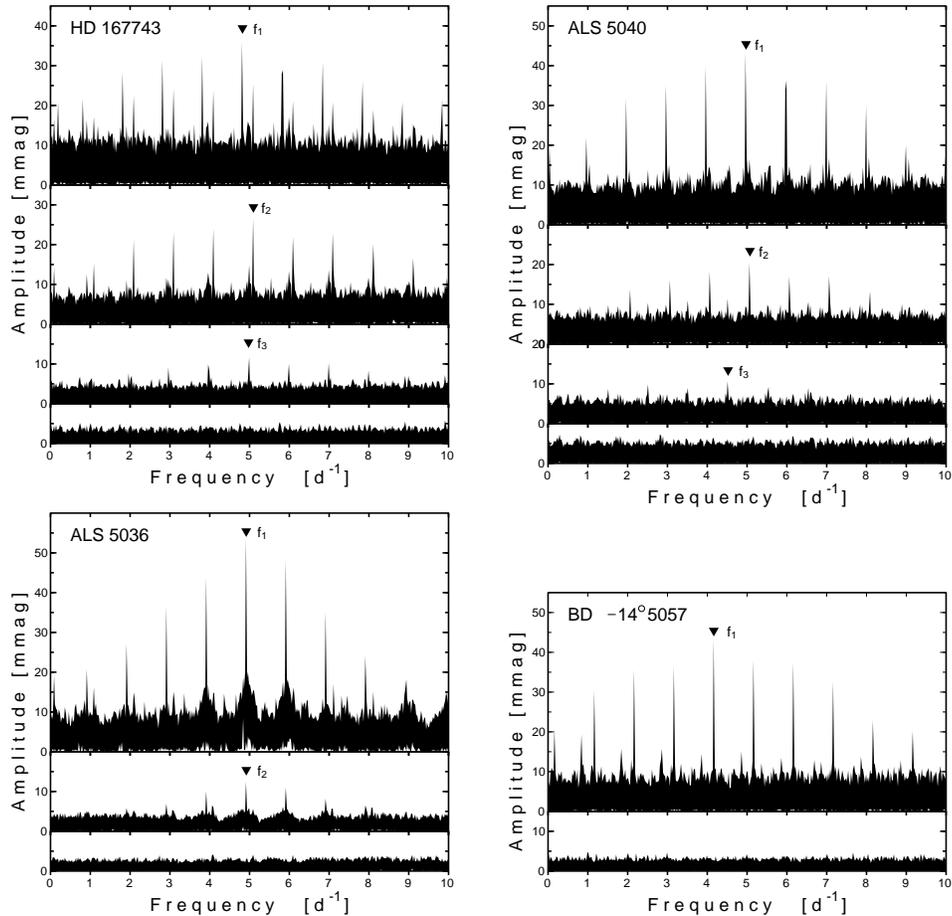}
\caption{\small The same as in Fig.~1, but for HD\,167743, ALS\,5036, ALS\,5040, and BD\,$-$14$^{\rm o}$5057.}
\label{four4}
\end{figure}

HD\,165582 = CPD\,$-$34$^{\rm o}$7600 = ALS\,4668.  It is certainly an early B-type star as confirmed by three
MK spectral classifications available in the literature:
B1\,II (Garrison {\it et al.}~1977), B1\,Ib (Houk 1982), and B0.5\,III (Clayton {\it et al.}~2000).  The star's
$UBV\beta$ photometry (KN77, Dachs {\it et al.}~1982, Schild {\it et al.}~1983) is consistent with the MK classification.  
The frequency spectrum,
shown in Fig.~3, is quite complicated.  
We detect four modes spread over a large range in frequency.  Owing to the severe
aliasing problem, it is possible that for some modes we did not extract the correct frequency.  
This applies especially to $f_2$.  We tried different alias frequencies for $f_2$, but the solutions other than that
shown in Table 2 always comprised more than five modes.  The solution we provide seems to be the best in the sense that 
it includes the frequencies of the highest peaks in the periodograms and consists of the smallest number of frequencies. 
However, the ambiguity remains, especially because $f_2 - f_4 \approx$ 4~d$^{-1}$.  The $f_3$ term suffers from the
same ambiguity despite
the fact that we detect its combination with $f_1$, namely $f_1+f_3$ (Fig.~3).  

HD\,167743 = BD\,$-$15$^{\rm o}$4909 = ALS\,9453.  Houk and Smith-Moore (1988) classified this
star as B2\,Ib.  No $UBV$ photometry is available, however.  The pulsational spectrum (Fig.~4) consists of
a triplet. The components, like those for HD\,152077, are not equidistant in frequency: $f_3 - f_1 \approx$ 0.152~d$^{-1}$,  
$f_2 - f_3 \approx$ 0.121~d$^{-1}$.

ALS\,5036.  No MK spectral type is available, but the $UBV$ photometry reported by Reed (1993) clearly indicates
that we are dealing with an early B-type star.  The frequency spectrum consists of two very close frequencies (Fig.~4), 
the beat period is of 
the order of 570~days.  This is very long, but a similar case is already known among $\beta$~Cephei stars:  the $f_3$ mode
in 16 Lac (Jerzykiewicz and Pigulski 1996).

ALS\,5040.  No MK spectral type is available.  However, Reed and Vance (1996) provide the $UBV$ photometry,
allowing us to conclude that it is a $\beta$~Cephei star.  The power spectrum (Fig.~4) reveals three independent
modes.

BD\,$-$14$^{\rm o}$5057 = ALS\,9636.  No MK spectral type is available, but the 
$UBV$ photometry of Lahulla and Hilton (1992) indicates an early B-type.  We detect a single frequency (Fig.~4).
\begin{figure}[!ht]
\epsfbox{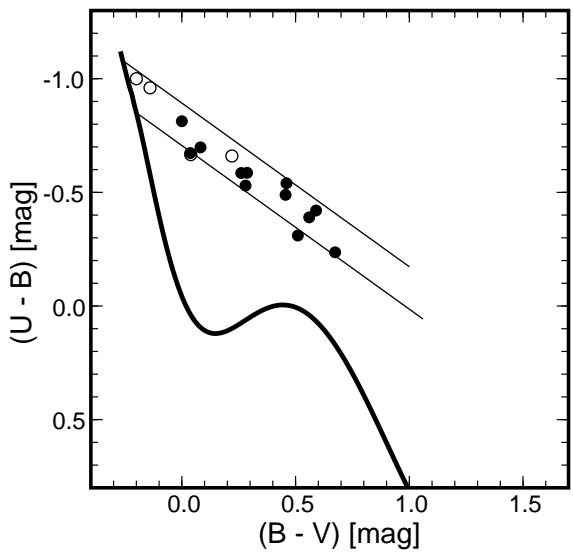}
\caption{\small The two-color diagram for all $\beta$~Cephei stars listed in Table 1 but HD\,328906 and HD\,167743.  
The thick 
continuous line is the unreddened main sequence.  The two thin lines are the reddening lines for a B0\,V star (upper)
and star B2.5\,V (lower) with the adopted $E(U-B)/E(B-V)$ = 0.72.  The four known $\beta$~Cephei stars listed in Table 1 are shown 
with open circles, the remaining twelve $\beta$~Cephei stars with the available $UBV$
photometry, with filled circles.}
\end{figure}

As a general comment to the $\beta$~Cephei stars that have no MK spectral types, we present 
Fig.~5, showing the location of the $\beta$~Cephei stars described in this paper in the two-color  
diagram. Only two stars, HD\,328906 and HD\,167743, out of 18 listed 
in Table 1, are not shown in the figure, because they lack $UBV$ photometry.
However, both have spectral types that allow us to classify them unambiguously as $\beta$~Cephei variables.
We see from Fig.~5 that all 16 stars, despite large range in reddening, fit well between the reddening 
lines for a B0 and B2.5\,V star.  This confirms that they are early B-type stars and supports their classification as
$\beta$~Cephei variables.

\vspace{0.5cm}
{\it 4.3.~Reclassified Stars}

\vspace{0.3cm}
As we indicated in the Introduction, 
out of 37 stars classified as BCEP in the ASAS-3 catalogue, only 12 were verified as $\beta$~Cephei stars.   
The remaining 25 stars were excluded because: (i) they have A or F MK spectral type indicating  
either $\delta$ Scuti or W\,UMa type (10 stars), (ii) the range of variability is too large ($\Delta V$ = 0.20--0.65~mag) and/or
the shape of the light curve indicated other type of variable (13 stars).  Of the large-amplitude variables classified as BCEP, 
nine are HADS, two are
RR~Lyrae stars of RRc type and two are W~UMa eclipsing binaries.  It is interesting to note that these stars are typically much fainter
than $\beta$~Cephei stars from Table 1, their $V$ magnitudes range between 11.4 and 13.3. 
  
There remain only two stars without MK types or $UBV$ photometry, ASAS\,202543+0948.0 and ASAS\,213518+1047.6, that have relatively
small amplitudes and the $V$ magnitudes of about 10.8.   For these two stars a chance that they are 
$\beta$~Cephei
stars remains.  We detect a single sinusoidal variation in ASAS\,202543+0948.0 and two frequencies
in ASAS\,213518+1047.6.  However, their dominant periods are equal to about 0.1~d, so that the $\delta$~Scuti 
classification seems to be more likely for them.  Their Galactic latitudes are equal to $-$15.8$^{\rm o}$ and $-$29.2$^{\rm o}$,
respectively, also indicating that
at least the second one is not a $\beta$~Cephei star.  
There are many stars similar to these two in the ASAS-3 catalogue, but they are usually classified as $\delta$~Scuti stars. Since 
neither the MK spectral type nor the $UBV$ photometry is available for them, we do not include 
the two, and the other stars with similar properties, in our list of $\beta$~Cephei stars.

\vspace{0.8cm}
\centerline{\bf 5. Discussion}

\vspace{0.5cm}
{\it 5.1.~Location in the Galaxy}

\vspace{0.3cm}
As we pointed out in the Introduction, the homogeinity of the ASAS-3 catalogue allows some general considerations.  First, we would 
like to comment on the distribution of the $\beta$~Cephei stars in the sky.  Their location in the Galactic coordinates is shown in 
Fig.~6.  As expected, they concentrate around the Galactic plane, with only four stars located at relatively high Galactic latitudes. 
Three are bright, and therefore nearby, while the fourth, HN\,Aqr = PHL\,346 (Waelkens and Rufener 1988, Hambly {\it et al.}~1996,
Dufton {\it et al.}~1998, Lynn {\it et al.}~2002), 
is the only known $\beta$~Cephei star located far from the Galactic plane.

Two other features in the distribution of the Galactic $\beta$~Cephei stars can be noted.  The first is the 
large asymmetry in the distribution: about 80\% of them 
are located in the southern sky.  This is partly due to the observational selection effects, but the dependence of the 
driving mechanism on metallicity, combined with decreasing metallicity at larger galactocentric distances, may also play an
important role (see Pigulski {\it et al.}~2002, Pigulski 2004).

The second feature seen in Fig.~6 is an apparent strip of bright $\beta$~Cephei stars between Galactic longitudes $l$ = 200$^{\rm o}$ 
and 360$^{\rm o}$, inclined with respect to the Galactic plane. It was already indicated by Lesh and Aizenman (1973) and explained as 
a part of the Gould Belt, a nearby (less than 1~kpc) apparent disk of OB stars surrounding 
the Sun and inclined by $\sim$20$^{\rm o}$ with respect to the Galactic plane (see, e.g., Stothers and Frogel 1974).
A part of the Gould Belt, at 300$^{\rm o} < l <$ 360$^{\rm o}$, is known as Sco-Cen OB association and forms an extended complex of
young stars at a distance of 0.1--0.2~kpc (e.g.~Sartori {\it et al.}~2003).  
\begin{figure}[!ht]
\epsfbox{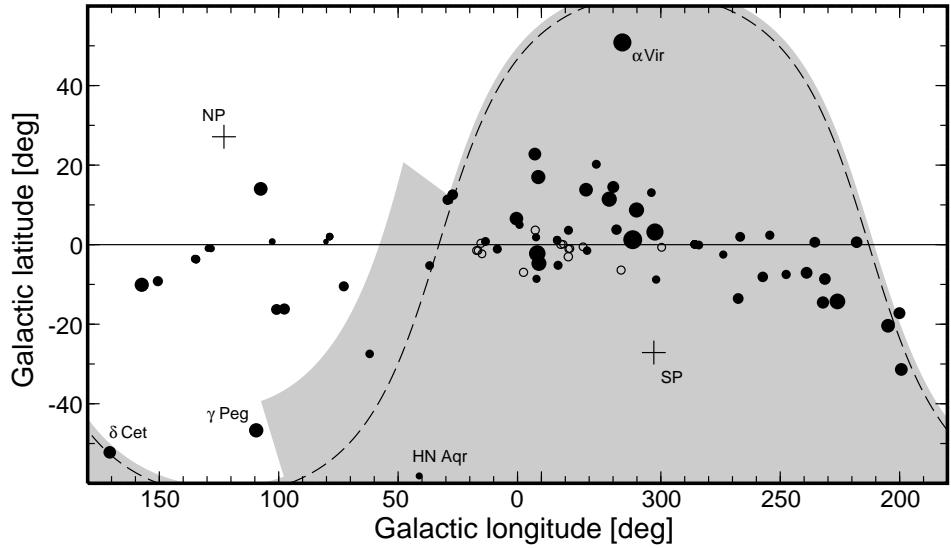}
\caption{\small Positions of known (filled circles) and new (open circles) $\beta$ Cephei stars in the Galactic coordinates.
Brighter stars are shown with larger symbols, those with the highest Galactic latitudes are labeled.  The grey
area shows the region of the sky covered by the ASAS-3 observations.  Dashed line shows the position of the celestial equator.  
Northern and 
southern celestial poles are indicated by plus signs and labeled `NP' and `SP', respectively.} 
\end{figure}
\begin{figure}[!t]
\epsfbox{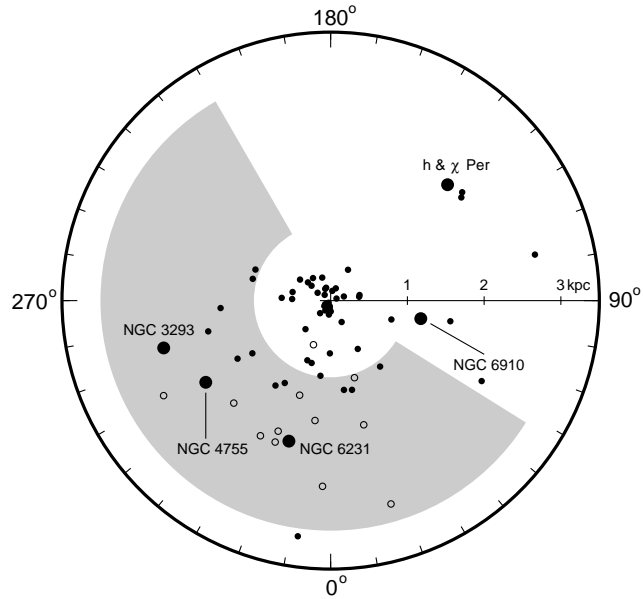}
\caption{\small The $\beta$~Cephei stars projected onto the Galactic plane. Filled and open circles denote known and new $\beta$~Cephei 
stars, respectively.  Open clusters with at least two $\beta$~Cephei stars are labeled and shown with 
larger symbols.  The position of the Sun is at the center of the plot.  Galactic longitudes are labeled at the circumference of the
large circle.  The grey area delimits roughly the region where $\beta$ Cephei
stars could be found from the ASAS-3 observations (see text for explanation).} 
\end{figure}

As the new $\beta$~Cephei stars are concerned, they are all located close to the Galactic plane.  Taking into account the
absolute magnitudes of $\beta$~Cephei stars, the range of their apparent 
magnitudes (9--11 mag, see Table 1) translates into the range of distances i.e.~of 1--3~kpc.  
Because the Galactic longitudes of the new $\beta$~Cephei stars cover $l$
between 300$^{\rm o}$ and 20$^{\rm o}$, we may conclude that they lie in the Sagittarius-Carina arm of the Galaxy.  
The fact that no new $\beta$~Cephei star was discovered in the interval 200$^{\rm o} < l <$ 300$^{\rm o}$ may also be explained.  
At these Galactic longitudes and the distances of 1--3~kpc we probe the areas where there is no pronounced spiral arm and therefore
no young population is present.  

This is even better illustrated in Fig.~7, where $\beta$~Cephei stars (except for HN~Aqr) are shown in the projection onto 
the Galactic plane.  The distances used in this figure were estimated using a simplified method of estimating absolute 
magnitude, $M_{\rm V}$.  First, the ($U-B$) and ($B-V$) colors were dereddened. Then, we assumed that $M_{\rm V}$ 
depends linearly on ($B-V$)$_0$.  For open clusters we used the distances from the literature.  The clusters containing
at least several $\beta$~Cephei stars are shown with large filled circles and labeled.

We see from Fig.~7 that, as pointed out above, the new $\beta$~Cephei stars populate mainly the Sagittarius-Carina arm of the Galaxy, 
where the three southern open clusters rich in $\beta$~Cephei stars, i.e., NGC\,3293, NGC\,4755, and NGC\,6231, are located too.

\vspace{0.5cm}
{\it 5.2.~The amplitudes}

\vspace{0.3cm}
As can be judged from Figs.~1--4, all $\beta$~Cephei stars found in the ASAS-3 catalogue have large amplitudes.  This can be
seen from Fig.~8 which shows the $V$-filter semi-amplitudes of the known $\beta$~Cephei stars (Stankov and Handler 2005) 
and the 14 new members described in this paper.  Among the previously known $\beta$~Cephei stars, BW Vul has indeed the
large amplitude.\footnote{Note 
that for BW Vul, Stankov and Handler (2005) report semi-amplitude, while for the other stars, the full amplitude.
This explains why BW~Vul does not 
stand out in their Figs.~4 and 8.}  The remaining stars have semi-amplitudes in $V$ below 40~mmag.  The new stars fill the gap, 
because their
main modes have semi-amplitudes larger than 35~mmag.  This is certailny due to the selection effect.  The detection threshold 
for all but two stars analyzed by us range between 5--10~mmag, the average value being 7~mmag.  Owing to the richness of 
the fields, the two stars in clusters are the only in our sample that have detection thresholds
exceeding 10~mmag (see Table 2).
We conclude therefore with
the statement which was also clear for the authors of the ASAS-3 catalogue: many $\beta$~Cephei stars (and other low-amplitude variables) 
can be found in the ASAS-3 data once the periodograms will be used as the search method, instead of the dispersion-magnitude diagram.

\begin{figure}[!ht]
\epsfbox{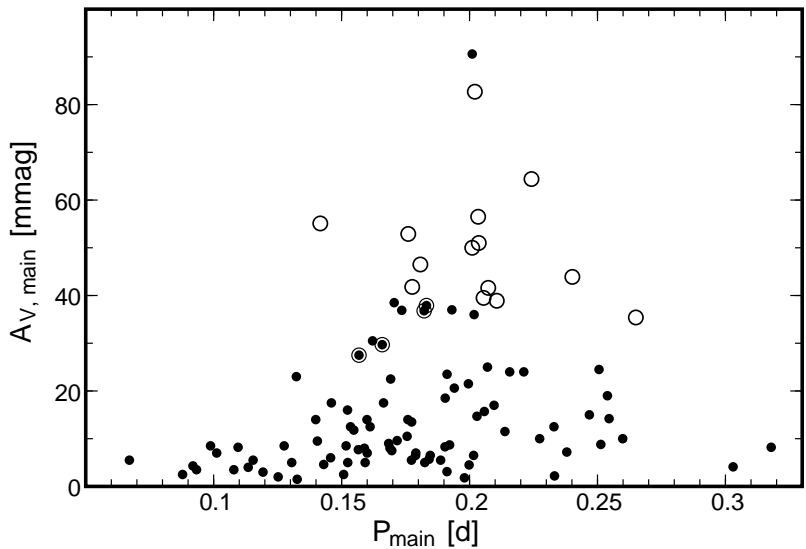}
\caption{\small $V$ semi-amplitudes of the dominating mode, $A_{\rm V, main}$, plotted against the period of this mode, $P_{\rm main}$, for
$\beta$~Cephei stars.  Dots denote stars from the catalogue of Stankov and Handler (2005), open circles, those
reported in this paper.  Encircled dots are four known $\beta$~Cephei stars in the ASAS-3 photometry (see Sect.~4.1).}
\label{amplit}
\end{figure}

\vspace{0.8cm}
\centerline{\bf 6. Conclusions}

\vspace{0.5cm}
The ASAS-3 photometry covering the whole southern sky turned out to be very efficient in finding new bright variable stars.  
Analysis of the ASAS-3 catalogue and the available literature led us to the conclusion that 14 stars from this catalogue 
can be now safely
classified as new $\beta$~Cephei stars.   All these stars have large amplitudes, many are multiperiodic.  They are
therefore excellent targets for follow-up spectroscopy and photometry.  As they are also bright ($V \approx$ 9--11~mag), 
this tasks can be performed by means of relatively small telescopes.  

\vspace{0.5cm}
{\bf Acknowledgements.} This work was supported by the KBN grant 1\,P03D\,016\,27.  This research has made use of the SIMBAD database,
operated at CDS, Strasbourg, France.  We wish to thank Prof.~M.\,Jerzykiewicz for helpful comments.

\vspace{0.5cm}
\centerline{REFERENCES}
\vspace{0.3cm}
{\small
\rea Aerts, C.~2000, {\it Astron.~Astrophys.}, {\bf 361}, 245.\par
\rea Clayton, G.C., Gordon, K.D., and Wolff, M.J.~2000, {\it Astrophys.~J.~Suppl.}, {\bf 129}, 147.\par
\rea Dachs, J., Kaiser, D., Nikolov, A., and Sherwood, W.A.~1982, {\it Astron.~Astrophys.~Suppl.}, {\bf 50}, 261.\par
\rea Dufton, P.L., Keenan, F.P., Kilkenny, D., {\it et al.}~1998, {\it M.N.R.A.S.}, {\bf 297}, 565.\par
\rea FitzGerald, M.P.~1987, {\it M.N.R.A.S.}, {\bf 229}, 227.\par
\rea FitzGerald, M.P., Jackson, P.D., and Moffat, A.F.J.~1977, {\it The Observatory}, {\bf 97}, 129.\par
\rea Forbes, D., and Short, S.~1996, {\it Astron.~J.}, {\bf 111}, 1609.\par
\rea Garrison, R.F., Hiltner, W.A., and Schild, R.E.~1977, {\it Astrophys.~J.~Suppl.}, {\bf 35}, 111.\par
\rea Hambly, N.C., Wood, K.D., Keenan, F.P., {\it et al.}~1996, {\it Astron.~Astrophys.}, {\bf 306}, 119.\par
\rea Handler, G., Shobbrook, R.R., Vuthela, F.F., Balona, L.A., Rodler, F., and Tshenye, T.~2003, {\it M.N.R.A.S.}, {\bf 341}, 1005.\par
\rea Haug, U.~1977, {\it The ESO Messenger}, {\bf 9}, 14.\par
\rea Haug, U.~1979, {\it Astron.~Astrophys.}, {\bf 80}, 119.\par
\rea Heynderickx, D., and Haug, U.~1994, {\it Astron.~Astrophys.~Suppl.}, {\bf 106}, 79.\par
\rea Hill, P.W., Kilkenny, D., and van Breda, I.G.~1974, {\it M.N.R.A.S.}, {\bf 168}, 451.\par
\rea Hogg, A.R.~1965, {\it Publ.~Astron.~Soc.~Pacific}, {\bf 77}, 440.\par
\rea Houk, N.~1978, {\it Michigan Spectral Survey, Ann Arbor, Dep.~Astron., Univ.~Michigan}, {\bf 2}.\par
\rea Houk, N.~1982, {\it Michigan Spectral Survey, Ann Arbor, Dep.~Astron., Univ.~Michigan}, {\bf 3}.\par
\rea Houk, N., and Cowley, A.P.~1975, {\it Michigan Spectral Survey, Ann Arbor, Dep.~Astron., Univ.~Mi\-chi\-gan}, {\bf 1}.\par
\rea Houk, N., and Smith-Moore, M.~1988, {\it Michigan Spectral Survey, Ann Arbor, Dep.~Astron., Univ.~Mi\-chi\-gan}, {\bf 4}.\par
\rea Jerzykiewicz, M., and Pigulski, A.~1996, {\it M.N.R.A.S.}, {\bf 282}, 853.\par
\rea Klare, G., and Neckel, T.~1977, {\it Astron.~Astrophys.~Suppl.}, {\bf 27}, 215 (KN77).\par
\rea Knude, J.~1992, {\it Astron.~Astrophys.~Suppl.}, {\bf 92}, 841.\par
\rea Lahulla, J.F., and Hilton, J.~1992, {\it Astron.~Astrophys.~Suppl.}, {\bf 94}, 265.\par
\rea Lesh, J.R., and Aizenman, M.L.~1973, {\it Astron.~Astrophys.}, {\bf 26}, 1.\par
\rea Lynn, B.B., Dufton, P.L., Keenan, F.P., {\it et al.}~2002, {\it M.N.R.A.S.}, {\bf 336}, 1287.\par
\rea Moffat, A.F.J., and Vogt, N.~1973, {\it Astron.~Astrophys.~Suppl.}, {\bf 10}, 135.\par
\rea Nesterov, V.V., Kuzmin, A.V., Ashimbaeva, N.T., Volchkov, A.A., R\"oser, S., and Bastian, U.~1995, {\it Astron.~Astrophys.~Suppl.},
    {\bf 110}, 367.\par
\rea Paczy\'nski, B.~1997, {\it Proc.~of the 12th IAP Colloquium:} ''Variable stars and the astrophysical 
    returns of microlensing surveys'', eds.~R.\,Ferlet, J.-P.\,Maillard, and B.\,Raban, p.~357.\par
\rea Paczy\'nski, B.~2000, {\it P.A.S.P.}, {\bf 112}, 1281.\par
\rea Paczy\'nski, B.~2001, {\it A.S.P.~Conf.~Ser.}, {\bf 246}, 45.\par
\rea Piatti, A.E., Clari\'a, J.J., and Bica, E.~2000, {\it Astron.~Astrophys.}, {\bf 360}, 529.\par
\rea Pigulski, A.~2004, {\it Comm.~in Asteroseismology}, {\bf 145}, 72.\par
\rea Pigulski, A., Kopacki, G., Ko{\l}aczkowski, Z., and Jerzykiewicz, M.~2002, {\it A.S.P.~Conf.~Ser.}, {\bf 259}, 146.\par
\rea Pojma\'nski, G.~1997, {\it Acta Astron.}, {\bf 47}, 467.\par
\rea Pojma\'nski, G.~2002, {\it Acta Astron.}, {\bf 52}, 397.\par
\rea Pojma\'nski, G.~2003, {\it Acta Astron.}, {\bf 53}, 341.\par
\rea Pojma\'nski, G., and Maciejewski, G.~2004, {\it Acta Astron.}, {\bf 54}, 153.\par
\rea Pojma\'nski, G., and Maciejewski, G.~2005, {\it Acta Astron.}, {\bf 55}, 97.\par
\rea Reed, B.C.~1993, {\it Astron.~J.}, {\bf 106}, 2291.\par
\rea Reed, B.C., and Vance, S.J.~1996, {\it Astron.~J.}, {\bf 112}, 2855.\par
\rea Roslund, C.~1964, {\it Arkiv f\"ur Astron.}, {\bf 3}, 357.\par
\rea Roslund, C.~1966, {\it Arkiv f\"ur Astron.}, {\bf 4}, 73.\par
\rea Rodr\'{\i}guez, E., and Breger, M.~2002, {\it A.S.P.~Conf.~Ser.}, {\bf 259}, 328.\par
\rea Sartori, M.J., L\'epine, J.R.D., and Dias, W.S.~2003, {\it Astron.~Astrophys.}, {\bf 404}, 913.\par
\rea Schild, R.E., Garrison, R.F., and Hiltner, W.A.~1983, {\it Astrophys.~J.~Suppl.}, {\bf 51}, 321.\par
\rea Stankov, A., and Handler, G.~2005, {\it Astrophys.~J.~Suppl.}, {\bf 158}, 193.\par
\rea Stothers, R., and Frogel, J.A.~1974, {\it Astron.~J.}, {\bf 79}, 456.\par
\rea Waelkens, C., and Rufener, F.~1988, {\it Astron.~Astrophys.}, {\bf 201}, L5.\par
\rea Waelkens, C., Aerts, C., Kestens, E., Grenon, M., and Eyer, L.~1998, {\it Astron.~Astrophys.}, {\bf 330}, 215.\par
\rea Walborn, N.R.~1971, {\it Astrophys.~J.~Suppl.}, {\bf 23}, 257.\par
\rea Whiteoak, J.B.~1963, {\it M.N.R.A.S.}, {\bf 125}, 105.\par

}

\end{document}